\documentclass[aps,prl, twocolumn,superscriptaddress]{revtex4-1}
\usepackage{graphicx}
\usepackage[hidelinks]{hyperref} 
\usepackage{bm}
\usepackage{amsmath}
\usepackage{bbold}
\usepackage{color}

\graphicspath{{/Users/shang/Documents/Paper/240125_MgCr2O4_theory/plots/}}

\begin{document}


\title{Dynamic Spin-Lattice Coupling and Statistical Interpretation for the Molecular-Like Excitations in Frustrated Pyrochlores} 

\renewcommand*{\thefootnote}{\arabic{footnote}}

\author{Shang Gao}
\email[]{sgao@ustc.edu.cn}
\affiliation{Department of Physics, University of Science and Technology of China, Hefei, Anhui 230026, People's Republic of China}

\date{\today}


\begin{abstract}
Emergent molecular-like excitations have been discovered in a series of pyrochlore antiferromagnets, yet their origins and relationships with the coexisting magnon excitations remain a puzzle. Here, by incorporating the dynamic spin-lattice coupling through the site-phonon model, we accomplish a unified description of the molecular and magnon excitations, which allows us to provide a statistical interpretation of the molecular-like modes. Our work also demonstrates a semiclassical approach to understand exotic spin dynamics in systems with non-bilinear spin interactions.
\end{abstract}

\maketitle

\textit{Introduction.} The exploration and understanding of emergent correlated states are among the central pursuits in condensed matter physics. This explains the excitements when hexamer-like molecular correlations were first discovered in the pyrochlore-lattice compound, ZnCr$_2$O$_4$~\cite{lee_local_2000, lee_emergent_2002}, which draws a close analogy to the self-organization phenomena widely observed in soft and hard condensed matters~\cite{hasegawa_self_1985, cross_pattern_1993, karsenti_self_2008}. Since then, a great variety of molecules~\cite{chung_statics_2005,matsuda_spin_2007, tomiyasu_molecular_2008, ji_spin_2009, matsuda_universal_2010, tomiyasu_Hg_2011, tomiyasu_Ge_2011, okamoto_breathing_2013, tomiyasu_spin_2014, tomiyasu_emergence_2013, gao_manifolds_2018, he_neutron_2021, gen_signatures_2023, kamazawa_magnetic_2003, dronova_three_2024}, including hexamers~\cite{lee_emergent_2002, tomiyasu_molecular_2008,he_neutron_2021}, heptamers~\cite{tomiyasu_molecular_2008,tomiyasu_emergence_2013, gao_manifolds_2018}, and their derivatives~\cite{tomiyasu_Hg_2011, dronova_three_2024}, have been proposed to phenomenologically describe the unusual spin dynamics in the frustrated pyrochlores. More recently, applications of phenomenological molecular models are extended to the triangular- and honeycomb-lattice compounds~\cite{yao_excitations_2022, gao_diffusive_2023, jin_magnetic_2023, chen_thermal_2024}, revealing surprising links of molecular correlations to topological spin textures~\cite{nagaosa_topological_2013, back_skyrmion_2020} and Kitaev quantum spin liquids~\cite{takagi_concept_2019, hermanns_physics_2018}.

The increasingly broad application of the phenomenological molecular models calls for urgent examinations of their physical origins. In ZnCr$_2$O$_4$ and the related pyrochlore antiferromagnets, although molecular-like excitations have been observed through neutron spectroscopy, there are many features in the spectra that cannot be explained by the free molecule models alone. One prominent discrepancy is the coexistence of dispersive spin wave excitations with the molecular-like modes~\cite{tomiyasu_molecular_2008,tomiyasu_emergence_2013,gao_manifolds_2018,nassar_pressure_2023,he_neutron_2021}, which often results in a separated treatment of the spin dynamics into two uncorrelated parts. In the absence of a unified description of the whole dynamic response function $S(\bm{Q}, \omega)$ as a function of wavevector transfer $\bm{Q}$ and energy $\omega$, it is challenging to conclude whether molecules are really existing or just effective simplifications of a more fundamental theory.

Recent studies of MgCr$_2$O$_4$, a model compound that is closely related to ZnCr$_2$O$_4$, cast doubts on the necessity of the molecular pictures in the paramagnetic regime~\cite{bai_magnetic_2019}. Through analysis of the $S(\bm{Q}, \omega)$ at high temperatures, it is found that the wavevector dependence of the scattering intensity, which was previously explained through free molecules~\cite{tomiyasu_molecular_2008,tomiyasu_emergence_2013,gao_manifolds_2018}, can be reproduced by a lattice Heisenberg model with couplings up to the third neighbors~\cite{conlon_absent_2010, bai_magnetic_2019}. However, in the ordered phase, the spin dynamics still remains puzzling, since the proposed Heisenberg model produces no molecular-like excitations below the LRO transition temperature, $T_N$.

The failure of the Heisenberg model at temperatures below $T_N$ may lie in the neglected lattice degree of freedom (DOF)~\cite{lee_local_2000, sushkov_probing_2005, chung_statics_2005, matsuda_spin_2007, kant_optical_2009, nilsen_complex_2015, rossi_negative_2019}. The importance of the lattice DOF is first established through the analysis of the 1/2-magnetization plateau that is often observed in this family of frustrated pyrochlores~\cite{ueda_field_2005, ueda_magnetic_2005, miyata_magnetic_2011, tsurkan_ultra_2017, gen_signatures_2023}. Assuming independent modulations in the bonding distances, a bond-phonon model has been proposed~\cite{tchernyshyov_order_2002, penc_half_2004}, which leads to effective biquadratic interactions over the neighboring spins and a consequent spin-Peierls transition with concurrent magnetic and structural transitions. On a  distorted lattice, spin wave calculations  indeed generate a resonant mode~\cite{tchernyshyov_order_2002}, yet the calculated structural factors, as discussed below, are rather different from those of the target molecules, suggesting the incompleteness of a statically distorted lattice model in describing the molecular-like dynamics.

Here we show that after considering the dynamic spin-lattice coupling, the site-phonon model~\cite{bergman_models_2006, wang_spin_2008, aoyama_spin_2016, aoyama_spin_2019}, a more physical spin-lattice model that assumes independent atomic vibrations, provides an elegant yet encompassing explanation for both the spin wave and molecular-like dynamics in frustrated pyrochlores. Our unified description establishes the importance of the dynamic spin-lattice coupling in determining the spin dynamics and also presents a statistical interpretation of the molecular models. Applications of this approach on both sides of $T_N$ are demonstrated for two model materials, MgCr$_2$O$_4$ and LiGaCr$_4$O$_8$, and shall be generalized to systems with either effective or intrinsic non-bilinear spin interactions.

\textit{Site-phonon model.} The site-phonon model on a regular pyrochlore lattice (see Fig.~\ref{fig:order}(a) in Appendix A for the crystal structure) can be described as

\begin{equation}
    {\cal H} = \sum_{\langle ij \rangle } J\left(|{\bm r}^{(0)}_{ij} + {\bm u}_i-{\bm u}_j| \right) {\bm S}_i \cdot {\bm S}_j + \frac{c}{2}\sum_i |{\bm u}_i|^2.
\label{eq:site_phonon}
\end{equation}
In this model, the exchange interaction $J$ between the nearest-neighboring spins $\bm{S}_i$ at sites $\bm r_i$ depends on their relative distance $\bm r_{ij} = \bm r_{ij}^{(0)} + \bm u_i - \bm u_j$, where the atomic displacements $\bm u_i$ from the original positions $\bm r_i^{(0)}$ are penalized by an elastic energy proportional to a positive constant $c$. Following the derivations in Refs.~\cite{bergman_models_2006, wang_spin_2008, aoyama_spin_2016, aoyama_spin_2019}, the spin-lattice coupling introduces two types of effective four-spin interactions, so that the Hamiltonian Eq.~(\ref{eq:site_phonon}) becomes $\mathcal{H} = \mathcal{H}_0+\mathcal{H}_{\textrm{SL}}$ with
\begin{align}
    {\cal H}_0 =& \bar{J}\sum_{\langle ij\rangle}\bm S_i \cdot \bm S_j \textrm{,} \qquad \\
    {\cal H}_{\rm SL} =&-\bar{J}\, b\,\sum_{\langle i,j \rangle } \left( {\bm S}_i \cdot {\bm S}_j \right)^2 \nonumber\\
    &-\frac{\bar{J}\, b}{2}\sum_i \sum_{j\neq k \in \mathrm{N}(i) }  {\bm{\hat{e}}}_{ij} \cdot {\bm{\hat{e}}}_{ik} \, \left( {\bm S}_i \cdot {\bm S}_j \right)\left( {\bm S}_i \cdot {\bm S}_k \right) , \qquad
\label{eq:effec}
\end{align}
where $\mathrm{N}(i)$ denotes the nearest neighbors of site $i$ and $b = (1/cJ)[(dJ/dr)|_{r=|\bm r_{ij}^{(0)}|}]^2$ is a positive parameter that measures the strength of the spin-lattice coupling. In the weak coupling regime of $b<0.25$ that is considered in our current work, the ground state exhibits a collinear magnetic order with a propagation vector $\bm{q}=(1,1,0)$ accompanied by a tetragonal lattice distortion~\cite{aoyama_spin_2016}, which is presented in Fig.~\ref{fig:order}(b) in Appendix A.



\textit{Numerical methods.} Staring from an initial spin configuration at time $t = 0$, the time evolution of classical spins in real space are solved using the equation of motion~\cite{moessner_low_1998, samarakoon_comp_2017, yan_half_2018, pohle_theory_2021}
\begin{equation}
    \frac{d\bm{S}_i}{dt} = \frac{i}{\hbar}[\bm S_i, \mathcal{H}] = \bm{H}_i\times\bm{S}_i\ \textrm{,}
\label{eq:md}
\end{equation}
where $\bm{H}_i$ is the molecular field at site $\bm{r}_i$.
For the site-phonon model in Eq.~(\ref{eq:effec}), the contribution of $\mathcal{H}_0$ to the molecular field is straightforward following the conventional Landau-Lifshitz-Gilbert equation~\cite{lakshmanan_fascinating_2011}
\begin{equation}
    \bm{H}^0_i = \bar{J}\sum_{j\in \mathrm{N}(i)}\bm{S}_j\ \textrm{.}
\label{eq:H0}
\end{equation}
The contribution of the four-spin Hamiltonian, $\mathcal{H}^{\textrm{SL}}$, can be treated using the random phase approximation (RPA). For the first term in Eq.~(\ref{eq:effec}), 
\begin{equation}
    \bm{H}^\textrm{SL(1)}_i \simeq -\bar{J}\, b\sum_{j\in \mathrm{N}(i)}\bar{S}_{ij}(t) \bm{S}_j\ \textrm{,}
\label{eq:HSL1}
\end{equation}
where $\bar{S}_{ij}(t) = \bm{S}_i(t)\cdot\bm{S}_j(t)$ with its time dependence explicitly shown. The second term in Eq.~(\ref{eq:effec}) involves three spins, therefore its contribution to the molecular field depends on the relative positions of the sites $i$, $j$, and $k$: 
\begin{align}
    \bm{H}^\textrm{SL(2)}_i \simeq &-\frac{\bar{J} \, b}{2} \sum_{j\neq k \in \mathrm{N}(i) }  {\bm{\hat{e}}}_{ij} \cdot {\bm{\hat{e}}}_{ik} \, \bar{S}_{ij}(t) {\bm S}_k \\
    &- \frac{\bar{J} \, b}{2} \sum_{\substack{j\in \mathrm{N}(i),\\ k \in \mathrm{N}(j), k\neq i}}  {\bm{\hat{e}}}_{ji} \cdot {\bm{\hat{e}}}_{jk} \, \bar{S}_{jk}(t) {\bm S}_j\ \textrm{.}
\label{eq:HSL2}
\end{align}

In a total molecular field $\bm H_i = \bm H_i^0 + \bm H_i^\textrm{SL(1)} + \bm H_i^\textrm{SL(2)}$, the equations of motion form a set of linear differential equations that can be solved numerically. Under such a treatment, the lattice DOF is considered implicitly~\cite{aoyama_spin_2016}, with the atomic displacements evolving in phase with the spin configurations so that the magnetoelastic energy is minimized at each time step.  Further details for our numerical simulations can be found in Appendix B.

\begin{figure*}[t]
    \includegraphics[width=1.0\textwidth]{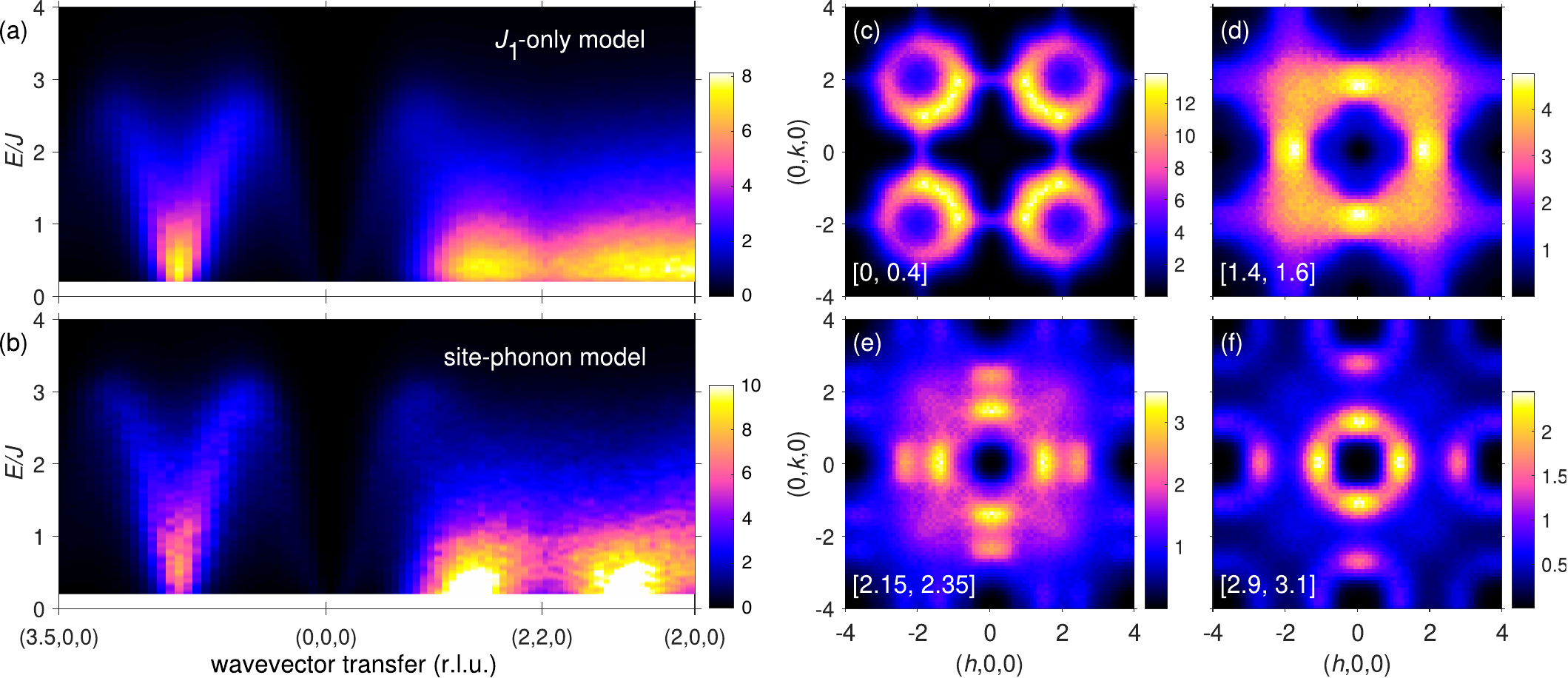}
    \caption{Dynamic response function, $S(\bm{Q}, \omega)$, on a regular pyrochlore lattice calculated for a $(10\times10\times10)$ supercell at $T = 0.2J$ using (a) the $J_1$-only Heisenberg model and (b) the site-phonon model with $b = 0.2$. Constant energy slices of the $S(\bm{Q}, \omega)$ for the site-phonon model are integrated in the energy ranges of (c) [0, 0.4], (d) [1.4 1.6], (e) [2.15, 2.35], and (f) [2.9, 3.1] in units of $J$. 
    \label{fig:highT}}
\end{figure*}

\begin{figure*}[t]
    \includegraphics[width=1.0\textwidth]{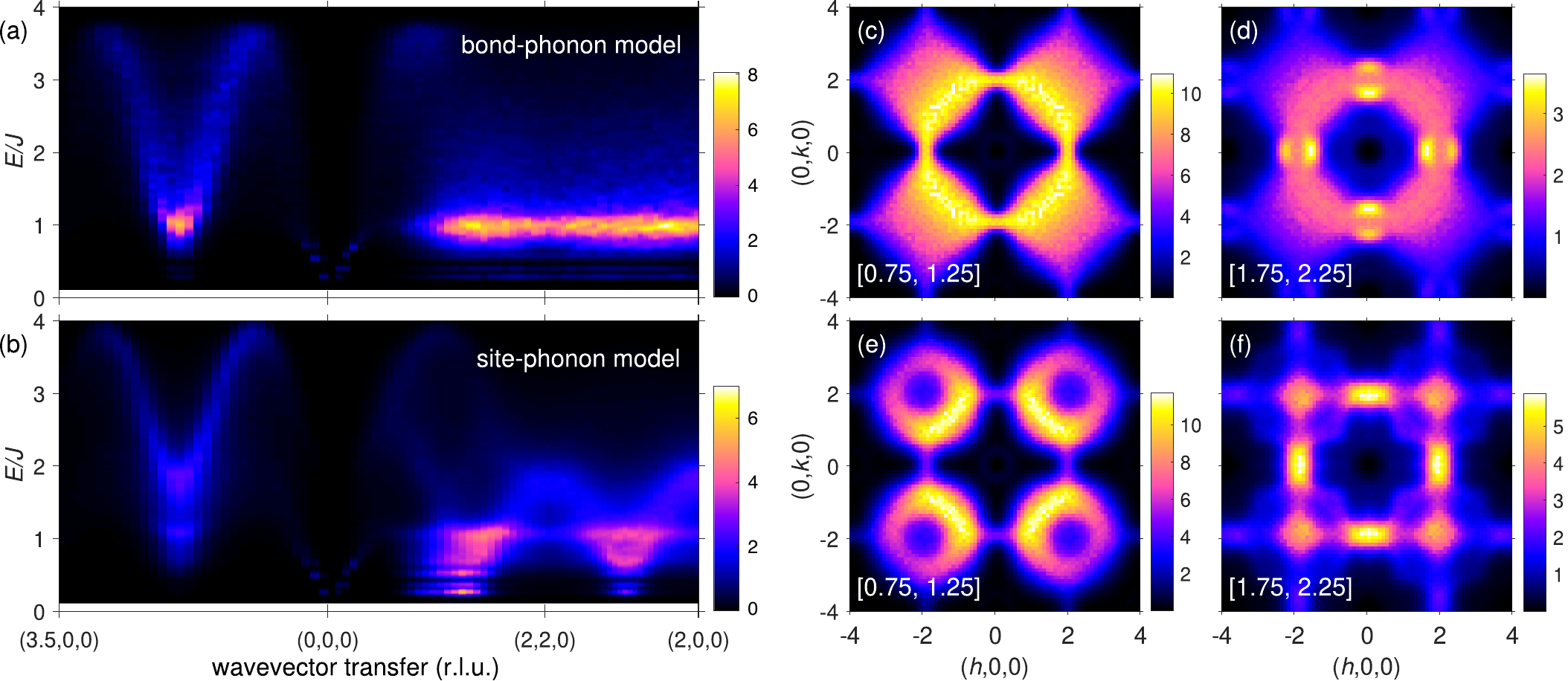}
    \caption{Comparison of the low-temperature $S(\bm{Q}, \omega)$ for (a) the bond-phonon model with $b = 0.13$ and (b) the site-phonon model with $b =0.2$ on a regular pyrochlore lattice. Calculations were performed at $T = 0.1J$ using a $(10\times10\times10)$ supercell. Constant energy slices of the $S(\bm{Q}, \omega)$ for the bond-phonon model are integrated in the energy ranges of (c) [0.75, 1.25] and (d) [1.75 2.25]. Similar slices for the site-phonon model are presented in panels (e) and (f).
    \label{fig:lowT}}
\end{figure*}
\textit{Molecular excitations above $T_N$.} Figures~\ref{fig:highT}(a) and (b) compare the dynamic response function, $S(\bm{Q}, \omega)$, for the $J_1$ Heisenberg model with only the nearest-neighboring bilinear couplings $J_1=J$ and the site-phonon model with $\bar{J}=J$ and $b=0.2$, calculated at temperatures above the LRO transition temperature $T_N$. For the latter model, $T_N$ is found to be at  $\sim 1.6J$. The symmetry lines in the plots follow those of the reported inelastic neutron scattering (INS) spectra for MgCr$_2$O$_4$~\cite{bai_magnetic_2019}, of which the spin correlations have been thoroughly characterized~\cite{tomiyasu_molecular_2008, tomiyasu_emergence_2013, gao_manifolds_2018, bai_magnetic_2019,nassar_pressure_2023}. To enable a direct comparison with the experimental data~\cite{bai_magnetic_2019}, the magnetic form factor of the Cr$^{3+}$ ions is applied over the calculated $S(\bm{Q}, \omega)$. For the convenience of comparison, Table~\ref{tab:compare} in Appendix C summarizes the previously reported experimental and calculated results that are related to each figures in our work. As revealed in Figs.~\ref{fig:highT}(a) and (b), the site-phonon model and the $J_1$-only model exhibit similar excitation spectra at energy transfers higher than $\sim 1.5J$. At lower energy transfers, the spin-lattice coupling strongly modulates the intensity distribution, leading to a spectra that is very similar to the experimental observations in MgCr$_2$O$_4$~\cite{bai_magnetic_2019}.

Figures~\ref{fig:highT}(c-f) present the constant energy slices of $S(\bm{Q}, \omega)$ for the site-phonon model. The quasielastic slice shown in Fig.~\ref{fig:highT}(c) reproduces the  experimental diffuse neutron scattering pattern~\cite{lee_emergent_2002, tomiyasu_emergence_2013, bai_magnetic_2019}, suggesting that the further-neighbor exchange couplings considered in the previous fits~\cite{bai_magnetic_2019} may at least partially arise from the spin-lattice couplings~\cite{bergman_models_2006, aoyama_spin_2016,aoyama_spin_2019}. Slices at higher energy transfers at $E \sim 1.5J$, $2.25J$, and $3J$ shown in Figs.~\ref{fig:highT}(d-f) also reproduce the experimental observations that were previously ascribed to the heptamer excitations~\cite{tomiyasu_emergence_2013}. Since the site-phonon model and the $J_1$-only model share similar $S(\bm{Q}, \omega)$ at $E\gtrsim 1.5J$, it can be concluded that the variations of the structural factors in MgCr$_2$O$_4$ at $E \sim 3J$ and $4J$, observed at temperatures both above and below $T_N$~\cite{tomiyasu_emergence_2013}, are better described by dispersive spin wave excitations.

\begin{figure}[t]
    \includegraphics[width=0.48\textwidth]{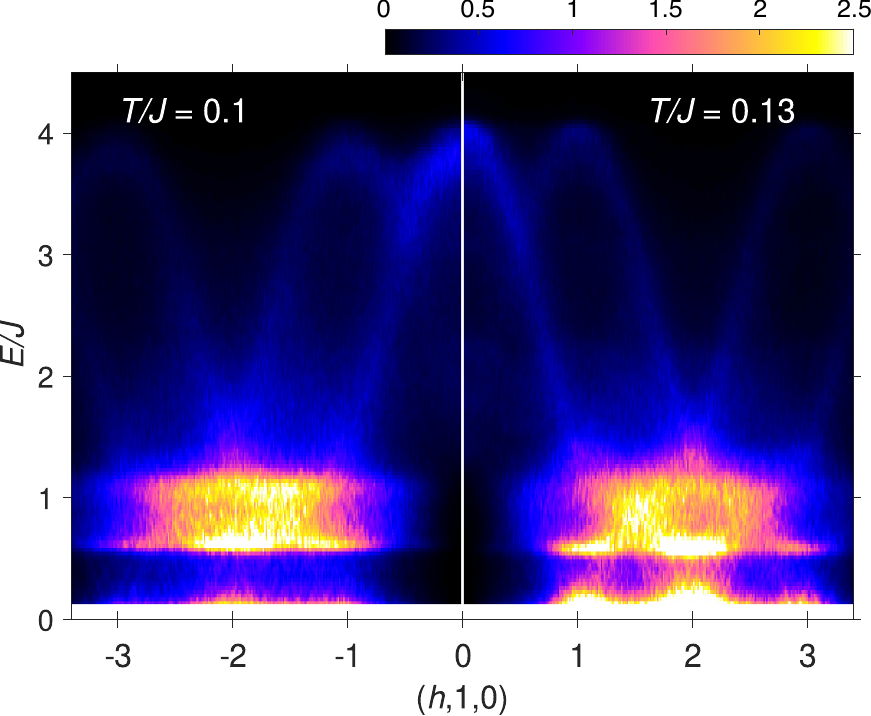}
    \caption{Dynamic response function $S(\bm{Q}, \omega)$ for the site-phonon model on a regular pyrochlore lattice along the ($\it{h}$,1,0) direction calculated using a $80\times 4\times 4$ supercell. Results in the left and right panels are calculated at $T=0.1J$ and $0.13J$, respectively, both being below $T_N \sim 0.16J$.
    \label{fig:split}}
\end{figure}

\textit{Molecular excitations below $T_N$.} As the failure of the bilinear Heisenberg model at temperatures below $T_N$ is already known~\cite{tchernyshyov_order_2002}, Fig.~\ref{fig:lowT} compares the low-temperature $S(\bm{Q}, \omega)$ of the site-phonon model to that of the bond-phonon model~\cite{tchernyshyov_order_2002, penc_half_2004}. Due to finite size effects, sharp flat modes appear in both spectra at $E\lesssim 0.5J$. These sharp modes can be suppressed by increasing the size of the supercell (see Fig.~\ref{fig:split} for calculations using a $80\times4\times4$ supercell). Similar to the linear spin wave calculations~\cite{tchernyshyov_order_2002}, a resonance-like mode is observed at $E \sim J$ for the bond-phonon model, which has been explained as local ring-like excitations on a distorted lattice. However, the flatness of the resonance mode and its constant energy slices shown in Figs.~\ref{fig:lowT}(c) and (d) deviate strongly from the experimental observations~\cite{tomiyasu_molecular_2008, tomiyasu_emergence_2013, gao_manifolds_2018,nassar_pressure_2023}. As a contrast, the site-phonon model, with the atomic dynamics implicitly considered, produces a resonance-like mode at $E \sim J$ that is highly dispersive, which is again in agreement with the experimental observations~\cite{gao_manifolds_2018,nassar_pressure_2023}. Due to this dispersion, the slices at $E\sim J$ and $2J$, shown in Figs.~\ref{fig:lowT}(e) and (f), respectively, are strongly modulated compared to those of the bond-phonon model. The intensity modulation reproduces all the details of the experimental observations that were previously ascribed to hexamer and heptamer excitations~\cite{tomiyasu_molecular_2008,tomiyasu_emergence_2013, gao_manifolds_2018}. 

Even the splitting of the resonance mode at $E \sim J$, which was recently established in experiments for MgCr$_2$O$_4$~\cite{gao_manifolds_2018,nassar_pressure_2023}, is reproduced in our calculations~\cite{tchernyshyov_order_2002, penc_half_2004}. Figure~\ref{fig:split} compares the $S(\bm{Q}, \omega)$ for the site-phonon model along the $(h,1,0)$ direction at $T = 0.1J$ and $0.13J$, both being below $T_N$. The shape of the resonance mode is well reproduced, including the tip at $E \sim 1.5J$ at wavevector transfer (2,1,0) and the band splitting at integer $h$. This splitting may arise from the domain effects as the tetragonal lattice distortion in the ordered phase leads to three crystallographic twins~\cite{aoyama_spin_2016}. Comparison between the $T = 0.1J$ and $0.13J$ data reveals that the magnon band softening at higher temperatures enlarges the hollow inside the split resonance modes around $(2,1,0)$. This observation is similar to the recent experimental discovery that the band splitting in MgCr$_2$O$_4$ becomes enhanced in high pressure~\cite{nassar_pressure_2023}, as the stiffness of the magnon bands may also be modulated by pressure.

\begin{figure}[t]
    \includegraphics[width=0.48\textwidth]{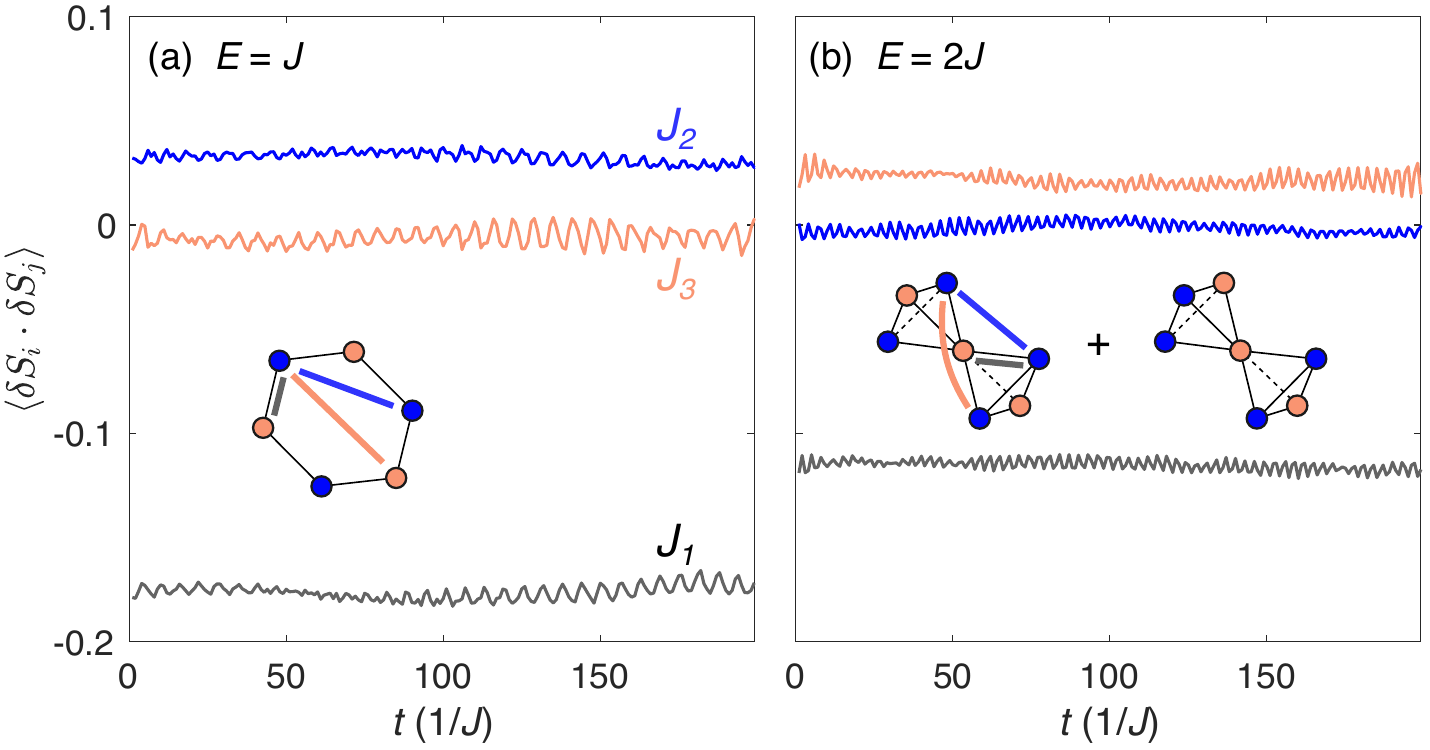}
    \caption{Time evolution of the correlations between the spin deviations $\langle \delta \bm{S}_i \cdot \delta \bm{S}_j \rangle$ at selected energies of $E \sim J$ (a) and $2J$ (b) for the site-phonon model with $b=0.2$. Correlations over the $J_1$, $J_2$, and $J_3$ bonds are shown in grey, blue, and orange, respectively. Insets are the corresponding molecular models with orange and blue spheres represent spin up and down, respectively. Typical paths for the $J_1$, $J_2$, and $J_3$ bonds are also indicated.
    \label{fig:stat}}
\end{figure}

Through a reverse Fourier transform, the time evolution of the spin configuration within a selected energy range can be recovered~\cite{pohle_theory_2021}. Following the proposal based on the spin-Peierls transition~\cite{tchernyshyov_order_2002}, at $E\sim J$, a stable and possibly regular lattice of hexamer-like pattern in real space is expected to emerge in the spin evolutions if ring-like excitations are stabilized by a \textit{static} lattice distortion. However, as evidenced in the video in the Supplemental Material, hexamer-like correlations appear only sporadically, and their relatively short lifetime, being consistent with their relatively broad energy width in the experimental spectra~\cite{gao_manifolds_2018, nassar_pressure_2023}, indicates the importance of the lattice dynamics that can create, propagate, and annihilate the molecular excitations. Following this scenario, a statistical interpretation of the molecular correlations can be established. Figure~\ref{fig:stat} plots the time evolution of the correlations between the spin deviations $\langle \delta \bm{S}_i \cdot \delta \bm{S}_j \rangle$ for different bonds averaged across a whole supercell. For the $E\sim J$ modes, the spin deviations are statistically ferromagnetic (antiferromagnetic) over the $J_2$ ($J_1$ and $J_3$) bonds, which is captured by the hexamer model shown in the inset of Fig.~\ref{fig:stat}(b)~\cite{tomiyasu_emergence_2013}. For the $E\sim 2J$ modes, correlations over the $J_2$ and $J_3$ bonds become statistically zero and ferromagnetic, respectively, which is also captured by the combination of two heptamers shown in the inset of Fig.~\ref{fig:stat}(b)~\cite{tomiyasu_emergence_2013}.

\textit{Breathing pyrochlore lattice.} Our dynamical treatment of the site-phonon model can be immediately generalized to the breathing pyrochlore lattice, on which the neighboring tetrahedra are of two different sizes and therefore two different exchange strengths $J$ and $J'$ are expected~\cite{aoyama_spin_2019, attila_dynamics_2024} (see Fig.~\ref{fig:order}(a) in Appendix A for the crystal structure). 
Figure~\ref{fig:breath} compares the powder-averaged dynamic response function of the site-phonon model with $J = 10.4$~meV and $J' = 6.2$~meV, together with a spin-lattice coupling of $b = 0.1$ on both types of tetrahedra. The strengths of the exchange couplings are chosen for a direct comparison with the experimental data for LiGaCr$_4$O$_8$, in which the coexistence of spin waves and molecular excitations have been experimentally observed but not explained~\cite{he_neutron_2021}. At $T=1$~K in the ordered phase, a resonance-like mode at $\sim 6$~meV, together with a wavevector dependence that was previously attributed to free hexamers~\cite{he_neutron_2021}, are reproduced. At $T = 50$~K in the paramagnetic phase, the resonance mode collapses to the elastic line, while its wavevector dependence stays almost unchanged, all in agreement with the experiments~\cite{he_neutron_2021}. 

\begin{figure}[t]
    \includegraphics[width=0.48\textwidth]{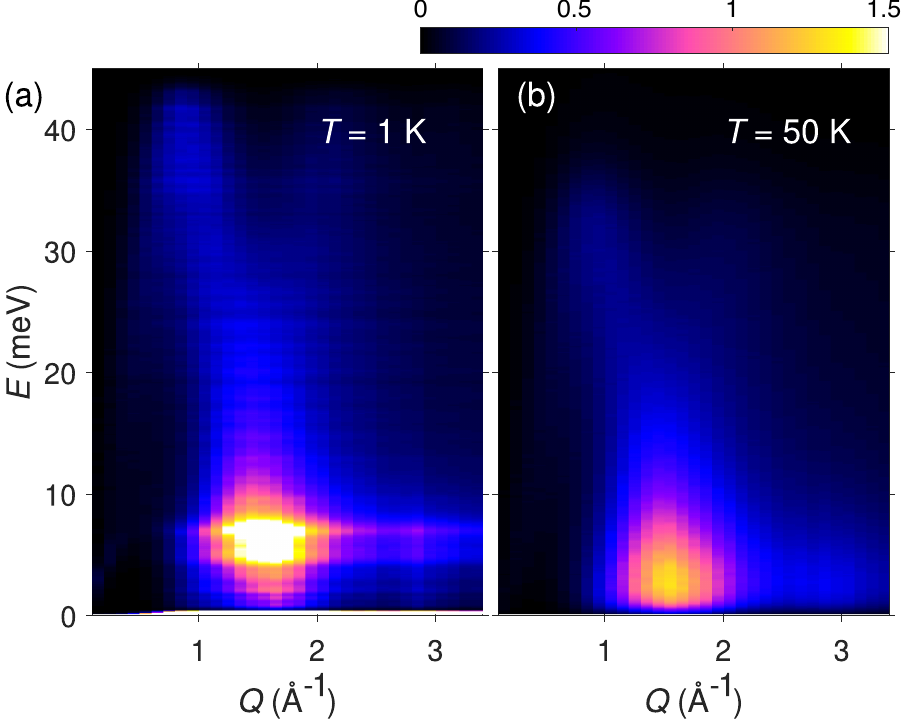}
    \caption{Powder-averaged $S(\bm{Q}, \omega)$ for the site-phonon model on a breathing pyrochlore lattice calculated for a $(10\times10\times10)$ supercell at $T = 1$~K (a) and 50~K (b). The coupling strengths are $J=10.4$~meV and $J'=6.2$~meV as proposed for LiGaCr$_4$O$_8$, with additional spin-lattice couplings of $b=0.1$ on both $J$ and $J'$ bonds. The magnetic form factor of the Cr$^{3+}$ ions is applied to compare with the experimental INS spectra.
    \label{fig:breath}}
\end{figure}

\textit{Conclusion and discussion.} By extending the site-phonon model to the dynamic regime, we successfully explain the coexistence of spin waves and molecular excitations in a series of frustrated pyrochlores as exemplified for two model compounds, MgCr$_2$O$_4$ and LiGaCr$_4$O$_8$. Our analysis reveals that the spin-Peierls transition alone is insufficient to explain the molecular-like dynamics in the frustrated pyrochlores. Instead, the dynamic spin-lattice coupling, which evolves the spin and lattice configurations simultaneously, is a necessary ingredient to fully understand the molecular-like dynamics.

Our dynamical approach to the site-phonon model shall be generalized to a large family of magnetoelastic materials, including the model triangular-lattice compounds NiGa$_2$S$_4$~\cite{nakatsuji_spin_2005, stock_neutron_2010, valentine_impact_2020} and $\alpha$-CaCr$_2$O$_4$~\cite{toth_magnetic_2012}, where the lattice DOF has been shown to be important yet the spin dynamics are not fully understood. On a broader scope, the semiclassical analysis demonstrated in our work may also be applied to systems where non-bilinear spin interactions arise not from spin-lattice coupling, but from multiple hopping of electrons~\cite{mila_origin_2000, wysocki_consistent_2011, paul_spin_2020, hoffmann_systematic_2020} or multipolar interactions~\cite{chen_exotic_2010}. As recently illustrated in the studies of topological spin textures~\cite{heinze_spontaneous_2011, grytsiuk_topo_2020} and Kitaev-type candidate materials~\cite{kruger_triple_2023, wang_effect_2023}, a better understanding of the origins and effects of the non-bilinear interactions can be crucial to explore and comprehend the rich spin correlated states in the strongly correlated electronic systems.





%


\begin{acknowledgments}
The author acknowledges helpful discussions with Oksana Zaharko, Tom Fennell, Christian R\"uegg, Joe Paddison, Xiaojian Bai, Yuan Li, Gang Chen, Erxi Feng, James Jun He, and Jyong-Hao Chen. The author thanks the generous host at the Interdisciplinary Center for Theoretical Study at USTC. Numerical calculations in this paper were partly performed on the supercomputing system in the Supercomputing Center of USTC. This project was funded by the National Science Foundation of China (NSFC) under the Grant No. 12374152.
\end{acknowledgments}

\appendix

\section{Appendix A: Lattice structure and the spin-lattice ground state}
\begin{figure}[h]
    \includegraphics[width=0.48\textwidth]{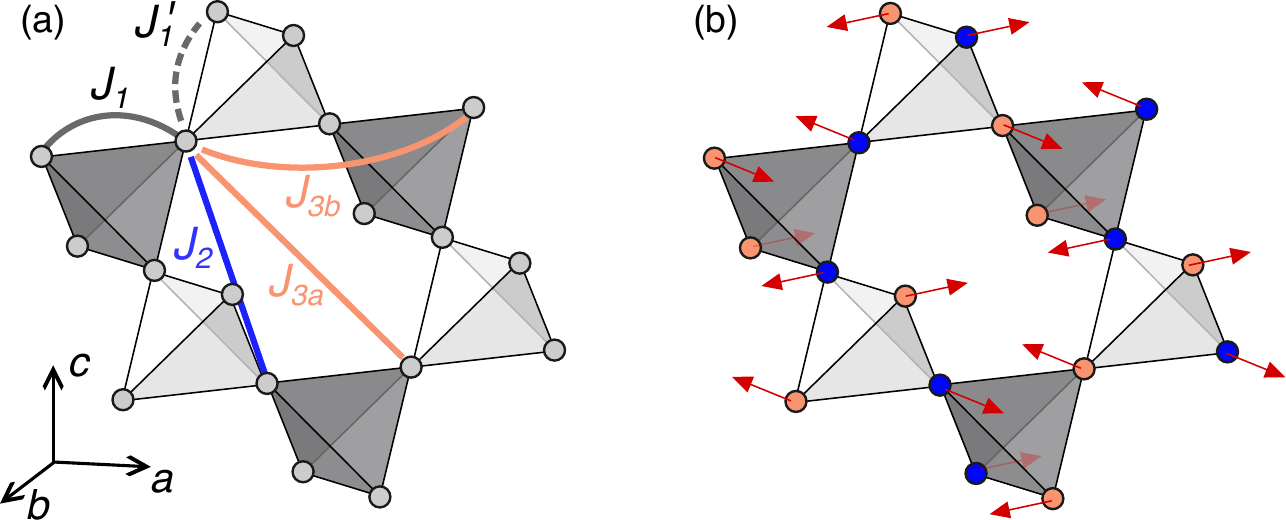}
    \caption{(a) Structure of the pyrochlore lattice. The bonding paths for couplings up to the third neigbors are indicated. In our calculations of $\langle \delta S_i \cdot \delta S_j \rangle$, both the $J_{3a}$ and $J_{3b}$ bonds are considered. On a breathing pyrochlore lattice, the different sizes of neighboring tetrahedra result alternating coupling strengths $J_1$ and $J_1'$. (b) Spin-lattice ground state of the site-phonon model in the weak coupling regime~\cite{aoyama_spin_2016}. Orange (blue) circles indicate spins pointing downwards (upwards). Red arrows indicate the atomic displacements in the $ab$ plane, which leads to a tetragonal lattice distortion.
    \label{fig:order}}
\end{figure}

Figure~\ref{fig:order}(a) presents the structure of the pyrochlore lattice that is composed of corner-sharing tetrahedra. Figure~\ref{fig:order}(b) shows the spin-lattice ground state for the site-phonon model in the weak coupling regime~\cite{aoyama_spin_2016}.

\section{Appendix B: Numerical details}

Unless otherwise specified, the initial spin configurations in our simulations were generated from classical Monte Carlo simulations on a $10\times10\times10$ supercell of conventional cubic cells after $1\times10^5\sim5\times10^5$ sweeps of single-spin Metropolis updates. Five overrelaxation sweeps were applied after each Monte Carlo sweep to reduce autocorrelation. For the dynamics calculations, spin configurations were evolved in a step size $\tau = 1/2J$ for a number of steps $N_t=400\sim800$ using Verner's “Most Efficient” 7/6 Runge-Kutta method~\cite{verner_numerically_2010}. The dynamic response function, $S(\bm{Q}, \omega)$, also called the spin excitation spectra, was then calculated through the fast Fourier transform (FFT)
\begin{equation}
S(\bm{Q}, \omega)=\frac{1}{\sqrt{N_t}N}\sum_{i,j}^{N}e^{i\bm{q}(\bm{r}_i-\bm{r}_j)}\sum_n^{N_t}e^{i\omega n\tau}\langle \bm{S}_i(0)\cdot \bm{S}_j(n\tau)\rangle\ \textrm{,}
\end{equation}
where $N$ is the total number of spins. A Parzen function with a halfwidth of $N_t/2$ was applied over the time evolution to reduce numerical artifacts~\cite{kuhner_dynamical_1999}.  For better statistics, $S(\bm{Q}, \omega)$ were averaged over $50\sim100$ independent runs. Following the discussions in Ref.~\cite{pohle_theory_2021}, a classical statistical factor of $\omega/(2k_\textrm{B}T)$ was applied over $S(\bm{Q}, \omega)$ to compare with the experimental data.

To generate the video in the Supplemental Material, the time evolution of the spin configuration was first transformed into the energy domain using the FFT. Then a Gaussian function with a full-width-half-maximum (FWHM) of $0.1J$ was applied to select the spin dynamics at $E\sim J$. After that, a reverse Fourier transform was applied to recover the time evolution of the spin deviations. In the video for the total spins that contains both the static and dynamic components, the static contribution was introduced through an additional Gaussian function with a FWHM of $0.1J$ at $E = 0$. A factor of 1/4 is applied to the static moment for better visibility of the spin dynamics.

The statistics of the correlations between the spin deviations $\langle \delta \bm{S}_i \cdot \delta \bm{S}_j \rangle$ shown in Fig.~\ref{fig:stat} was calculated by similar reverse Fourier transform methods. As shown in Fig.~\ref{fig:order}(a) in Appendix A, there are two types of third-neighbor bonds on a regular pyrochlore lattice, denoted as $J_{3a}$ and $J_{3b}$. In our simulations, these two bonds exhibit similar time dependence and therefore are not differentiated in Fig.~\ref{fig:stat}.

\section{Appendix C: Comparisons with the previous experimental data and molecular models}
For the convenience of comparison, Table~\ref{tab:compare} lists the previously reported experimental and calculated molecular spectra that are related to each figures in our work.

\begin{table}[h]
    \caption{\label{tab:compare} Previously reported experimental and calculated spectra that are related to each figures in our work.}
    \begin{tabular}{l l l}
    \hline 
    Our Work \quad &  \quad Experiments \quad &  \quad Molecular models \quad \\ \hline
    Fig.~1(b)  &  \quad Fig.~2(a) in \cite{bai_magnetic_2019}    &   \quad Fig.~2(a) in \cite{bai_magnetic_2019}   \\ 
    Fig.~1(c)  &  \quad Fig.~3 in \cite{lee_emergent_2002}    &   \quad Fig.~3 in \cite{lee_emergent_2002}   \\
    &  \quad Fig.~1(d) in \cite{bai_magnetic_2019}    &   \quad -    \\
    Figs.~1(d)-(f) &  \quad Fig.~1(d) in \cite{tomiyasu_emergence_2013}    &   \quad Fig.~1(f) in \cite{tomiyasu_emergence_2013}   \\
    Fig.~1(e) &  \quad Fig.~5(d) in \cite{gao_manifolds_2018}    &   \quad Fig.~10 in \cite{gao_manifolds_2018}   \\
    Fig.~2(a) &  \quad -    &   \quad Fig.~5 in \cite{tchernyshyov_order_2002}   \\
    Fig.~2(b) &  \quad Fig.~6(b) in \cite{gao_manifolds_2018}    &   \quad -   \\
    Figs.~2(e),(f) &  \quad Figs.~5(b),(c) in \cite{gao_manifolds_2018}    &   \quad Fig.~10 in \cite{gao_manifolds_2018}   \\
    &  \quad Figs.~2(a),(c) in \cite{tomiyasu_molecular_2008}    &   \quad Figs.~2(e),(g) in \cite{tomiyasu_molecular_2008}   \\
    &  \quad Fig.~1(e) in \cite{tomiyasu_emergence_2013}    &   \quad Fig.~1(f) in \cite{tomiyasu_emergence_2013}   \\
    Fig.~3 &  \quad Fig.~7(a) in \cite{gao_manifolds_2018}    &   \quad -   \\
    &  \quad Fig.~5 in \cite{nassar_pressure_2023}    &   \quad -   \\
    Fig.~4(a) &  \quad Fig.~3(a) in \cite{he_neutron_2021}    &   \quad Fig.~3(d) in \cite{he_neutron_2021}  \\
    Fig.~4(b) &  \quad Fig.~1(a) in \cite{he_neutron_2021}    &   \quad -   \\
    \hline
\end{tabular}
\end{table}

%
    
\end{document}